\documentclass[usenatbib,twocolumn,trackchanges]{aastex7}
\usepackage{natbib}

\newcommand{\av}{\ensuremath{A(V)}}
\newcommand{\rv}{\ensuremath{R(V)}}
\newcommand{\ebv}{\ensuremath{E(B-V)}}
\newcommand{\nhi}{\ensuremath{N(HI)}}
\newcommand{\nhiebv}{\ensuremath{N(HI)/E(B-V)}}
\newcommand{\nhiav}{\ensuremath{N(HI)/A(V)}}

\newcommand{\elvebv}{\ensuremath{E(\lambda - V)/E(B - V)}}

\shorttitle{Dust in M31}
\shortauthors{Clayton et al.}

\begin{document}

\title{Nature or Nurture: LMC-like Dust in the Solar Metallicity Galaxy M31}


\author[0000-0002-0141-7436]{Geoffrey C. Clayton}
\affiliation{Space Science Institute,
4765 Walnut St., Suite B,
Boulder, CO 80301, USA}
\email[show]{
gclayton@spacescience.org}
\affiliation{Department of Physics \& Astronomy, Louisiana State University, Baton Rouge, LA 70803, USA}
\affiliation{Maria Mitchell Association, 4 Vestal St., Nantucket, MA 02554, USA}

\author[0000-0002-9912-6046]{Petia Yanchulova Merica-Jones}
\affiliation{Space Telescope Science Institute, 3700 San Martin Drive, Baltimore, MD 21218, USA}
\affiliation{University of Sofia, Faculty of Physics, 5 James Bourchier Blvd., 1164 Sofia, Bulgaria}
\email{petiay@gmail.com}

\author[0000-0001-5340-6774]{Karl D. Gordon}
\affiliation{Space Telescope Science Institute, 3700 San Martin Drive, Baltimore, MD 21218, USA}
\affiliation{Sterrenkundig Observatorium, Universiteit Gent, Krijgslaan 281 S9, B-9000 Gent, Belgium}
\email{kgordon@stsci.edu}

 \author[0000-0001-9462-5543]{Marjorie Decleir}
\affiliation{European Space Agency (ESA), ESA Office, Space Telescope Science Institute (STScI), 3700 San Martin Drive, Baltimore, MD 21218, USA}
\altaffiliation{ESA Research Fellow}
\email{mdecleir@stsci.edu}

\author[0000-0002-7743-8129]{Claire E. Murray}
\affiliation{Space Telescope Science Institute, 3700 San Martin Drive, Baltimore, MD 21218, USA}
\affiliation{Department of Physics \& Astronomy, The Johns Hopkins University, 3400 N. Charles St., Baltimore, MD 21218, USA}
\email{cmurray1@stsci.edu}

\author[0000-0001-9806-0551]{Ralph Bohlin}
\affiliation{Space Telescope Science Institute, 3700 San Martin Drive, Baltimore, MD 21218, USA}
\email{bohlin@stsci.edu}

\author[0000-0001-9806-0551]{Luciana Bianchi}
\affiliation{Department of Physics \& Astronomy, The Johns Hopkins University, 3400 N. Charles St., Baltimore, MD 21218, USA}
\email{lbianch1@jhu.edu}

\author[0000-0001-6563-7828]{Philip Massey}
\affiliation{Lowell Observatory, 1400 W Mars Hill Road, Flagstaff, AZ 86001, USA}
\affiliation{Department of Astronomy and Planetary Science, Northern Arizona University, Flagstaff, AZ 86011-6010, USA}
\email{massey@lowell.edu}

\author[0000-0002-1127-8329]{Michael J. Wolff}
\affiliation{Space Science Institute,
4765 Walnut St., Suite B,
Boulder, CO 80301, USA}
\email{mjwolff@SpaceScience.org}

\begin{abstract}
Using the {\it Hubble Space Telescope}/Space Telescope Imaging Spectrograph, ultraviolet (UV) extinction curves have been measured in M31 along thirteen new sightlines, increasing the M31 sample to seventeen.
This sample covers a wide area of M31 having galactocentric distances of 5 to 16 kpc, enabling the analysis of UV extinction curve variations over a large region of an external galaxy similar to the Milky Way with global galactic characteristics such as metallicity for the first time. 
No correlation is found between the extinction parameters and galactocentric distance which might be expected if there is a radial metallicity gradient in M31.
Most of the new UV extinction curves presented here are significantly different from the average extinction curves of the Milky Way, LMC, and SMC, but the average M31 extinction curve is similar to the average extinction curve in the 30-Dor region of the LMC. 
The wide range of extinction curves seen in each individual Local Group galaxy suggests that global galactic properties such as metallicity may be less important than the local environmental conditions such as density, UV radiation field, and shocks along each sightline.
The combined behavior of the Milky Way, LMC, SMC, and now M31 UV extinction curves supports the idea that there is a family of curves in the Local Group with overlapping dust grain properties between different galaxies.
\end{abstract}


\section{Introduction}

Interstellar dust is the dominant opacity source for photons with energies between the microwave and the ionization edge of hydrogen.
Extinction curve measurements provide critical constraints on the size, composition, and shape of dust grains.
In addition, such curves empirically allow for the effects of dust extinction to be accounted for in observations of many astrophysical objects.
Understanding dust is important on many scales including entire galaxies, as a large fraction of the ultraviolet (UV) stellar photons passing through the ISM in galaxies is absorbed and re-emitted in the IR.
Hence, the energy budget of a galaxy cannot be accurately modeled using radiative transfer codes without a detailed knowledge of the characteristics of this absorbing medium. 

Typically, the UV extinction characteristics can only be mapped out in galaxies where individual stars are bright enough for UV spectroscopy.
Because of this limitation, almost all of the studies of UV extinction have been limited to only three galaxies in the Local Group, the Milky Way (MW) \citep[e.g.,][]{1989ApJ...345..245C,1990ApJS...72..163F,2000ApJS..129..147C,2004ApJ...616..912V}, the Large Magellanic Cloud (LMC) \citep[e.g.,][]{1985ApJ...288..558C,1985ApJ...299..219F,1999ApJ...515..128M,2003ApJ...594..279G}, and the Small Magellanic Cloud (SMC) \citep[e.g.,][]{1998ApJ...500..816G,2003ApJ...594..279G,2012A&A...541A..54M,2024ApJ...970...51G}. 
Pathfinder studies of a few sightlines in M31 \citep{1996ApJ...471..203B,2015ApJ...815...14C} have shown M31 to have similar UV extinction curves to those commonly seen in the Milky  Way and LMC.

Combined, these studies make up the unique database of UV extinction curves for Local Group galaxies.
The goal of these works and our new analysis presented in this paper is to map the properties of interstellar dust across a sample of Local Group galaxies with different global characteristics such as metallicity and star formation activity. 
The MW and M31 are metal rich \citep[e.g.,][]{2014ApJ...780..172D}. The LMC and, particularly, the SMC have sub-solar metallicities \citep[][and references therein]{2016ARA&A..54..363D}. The SMC metallicity is about 20\% Solar and the LMC metallicity is about 50\% Solar.

\citet[][hereafter CCM89]{1989ApJ...345..245C} found that the wavelength dependence of most observed extinction curves in the MW could be described by one parameter, $\rv\ = \av/\ebv$, the ratio of total to selective extinction which is a measure of the average dust grain size.
This work was extended to over 400 sightlines in the MW of which only 4 differed significantly from the R(V) dependent relationship in CCM89 \citep{2004ApJ...616..912V}. 
It was recognized very early in UV studies that the dust extinction properties in the LMC and SMC are different from the MW and from each other \citep[e.g.,][]{1984A&A...132..389P,1985ApJ...288..558C}.
It has become increasingly apparent that ``standard'' MW type dust, as characterized by the CCM89 extinction relation, does not provide the best fit to many extragalactic SEDs.
For instance, radiative transfer models of many starburst galaxies seem to imply dust properties similar to those associated with the SMC \citep{1997ApJ...487..625G,1998ApJ...500..816G}.
It is not surprising that extinction in other galaxies has different properties than the MW given the limited dust environments probed by the CCM89 and later studies (i.e., within 1 kpc of the Sun). 
So far, the variations of interstellar dust properties within a galaxy have only been mapped spatially over the entire galaxy in the Magellanic Clouds, both of which show significant extinction variations with spatial position \citep{2003ApJ...594..279G,2024ApJ...970...51G}. 

The first attempt to study the UV extinction of dust in M31 was almost 30 years ago.
A very small sample of OB stars in M31 was observed using {\it Hubble Space Telescope}/Faint Object Spectrograph (HST/FOS) but the data had very low S/N \citep{1996ApJ...471..203B}.
An average M31 extinction curve, derived from only three sightlines, has an overall wavelength dependence similar to that of the average MW extinction curve, but potentially a weaker 2175~\AA\ bump (significance of $\sim$1$\sigma$).
\citet{2015ApJ...815...14C} obtained new low-resolution UV spectra of a small sample of reddened OB stars in M31 with the {\it Hubble Space Telescope}/Space Telescope Imaging Spectrograph (STIS).
Extinction curves were constructed for four additional sightlines in M31 paired with closely matching stellar atmosphere models.
Direct measurements of \nhi\ were also made using the Ly$\alpha$ absorption lines enabling gas-to-dust ratios to be calculated.
The measured extinction curves are similar to those seen in the MW and LMC.
A dust grain model was used to investigate the dust composition and size distribution for the sightlines observed in this program, finding that the extinction curves can be produced with the available carbon and silicon abundances, if the metallicity is super-solar.

In this paper, we report the extinction properties in M31 along 13 additional lines of sight providing the opportunity to look for regional variations in the dust extinction properties across M31 for the first time.

\section{Observations and Analysis}

\subsection{Sample}

\begin{deluxetable*}{lllllllrl}
\label{tb_targets}
\tablecaption{Targets}
\tablehead{\colhead{Star\tablenotemark{a}} & \colhead{$\alpha_{2000.0}$} & \colhead{$\delta_{2000.0}$} & \colhead{V} & \colhead{B-V} & \colhead{SpT\tablenotemark{b}} & \colhead{E(B-V)} & \colhead{D\tablenotemark{c}} & \colhead{ID} }
\startdata
J004354.05+412626.0 & 00 43 54.05 &+41 26 26.0& 19.04 &  0.11 &  O7.5III  &  0.42& 5.62&   e1\\
J004413.84+414903.9 & 00 44 13.84& +41 49  3.9 &18.85& 0.08& O9.7Ia & 0.35&11.01& e2\\ 
J004420.52+411751.1  &00 44 20.52 &+41 17 51.1 &18.63  &0.15 &B0.5Ia &  0.37   &14.22&   e3\\
J004427.47+415150.0&  00 44 27.47 &+41 51 50.0 &19.15& 0.24  &B1.5Ia &  0.42  &11.45&    e4\\
J004431.65+413612.4 & 00 44 31.65 &+41 36 12.4& 16.79& 0.25& B2.5Ia   &0.40    &7.47&  e5\\
J004438.75+415553.6 & 00 44 38.75& +41 55 53.6 &17.25 & 0.24 &  B2.5Ia  & 0.39  &12.80&    e6\\
J004454.37+412823.9 & 00 44 54.37 &+41 28 23.9 &18.49  &0.34 &B1Ie&0.53&13.62&e7 \\
J004511.82+415025.3&  00 45 11.82 &+41 50 25.3 &18.23 & 0.35&  B2Ia &   0.52   &10.06&    e8\\
J004511.85+413712.9 & 00 45 11.85 &+41 37 12.9 &18.40  &0.10  & O9 III(f)&  0.41&12.15& e9\\
J004539.00+415439.0 & 00 45 39.00& +41 54 39.0 &19.50 & 0.32& B2.5V&0.47&11.74&e12\\
J004539.70+415054.8 & 00 45 39.70& +41 50 54.8 &19.36 & 0.35 & B2.5I &   0.50  &11.90&    e13\\
J004543.48+414513.7 & 00 45 43.48 &+41 45 13.7 &19.08 &   0.39&   B2I  &   0.56   &13.51&    e14\\
J004546.81+415431.7&  00 45 46.81& +41 54 31.7 &18.80  & 0.21 & B2.5I   &  0.36  &12.24&   e15\\
		\hline
  \hline
J003944.71+402056.2\tablenotemark{d}& 00 39 44.70& +40 20 56.1& 18.2 &0.15& O9.7Ib &0.42&16.36&e17\\
J003958.22+402329.0\tablenotemark{d}& 00 39 58.21 &+40 23 29.0 &18.97& 0.09& B0.7Ia &0.30&15.87&e18\\
J004034.61+404326.1\tablenotemark{d}& 00 40 34.61 &+40 43 26.1& 18.67 &0.15& B1Ia &0.34&9.32&e22\\
J004412.17+413324.2\tablenotemark{d}& 00 44 12.16& +41 33 24.1& 17.33& 0.34& B2.5Ia &0.49&5.99&e24\\
\enddata
\tablenotetext{a}{The star names are based on their celestial (J2000.0) positions \citep{2006AJ....131.2478M}. They can be searched on SIMBAD by placing LGGS in front of the star name in column one.}
\tablenotetext{b}{Spectral types from the literature \citep{1995AJ....110.2715M,2016AJ....152...62M}.}
\tablenotetext{c}{M31 Galactocentric distance in kpc. Calculated using \url{https://github.com/PBarmby/pb_utils/blob/master/gal_radii_pb.py}}
\tablenotetext{d}{This target was previously analyzed in \citet{2015ApJ...815...14C}.}
\end{deluxetable*}

\begin{figure}[tbp]
\centering
\includegraphics[width=3.2in]{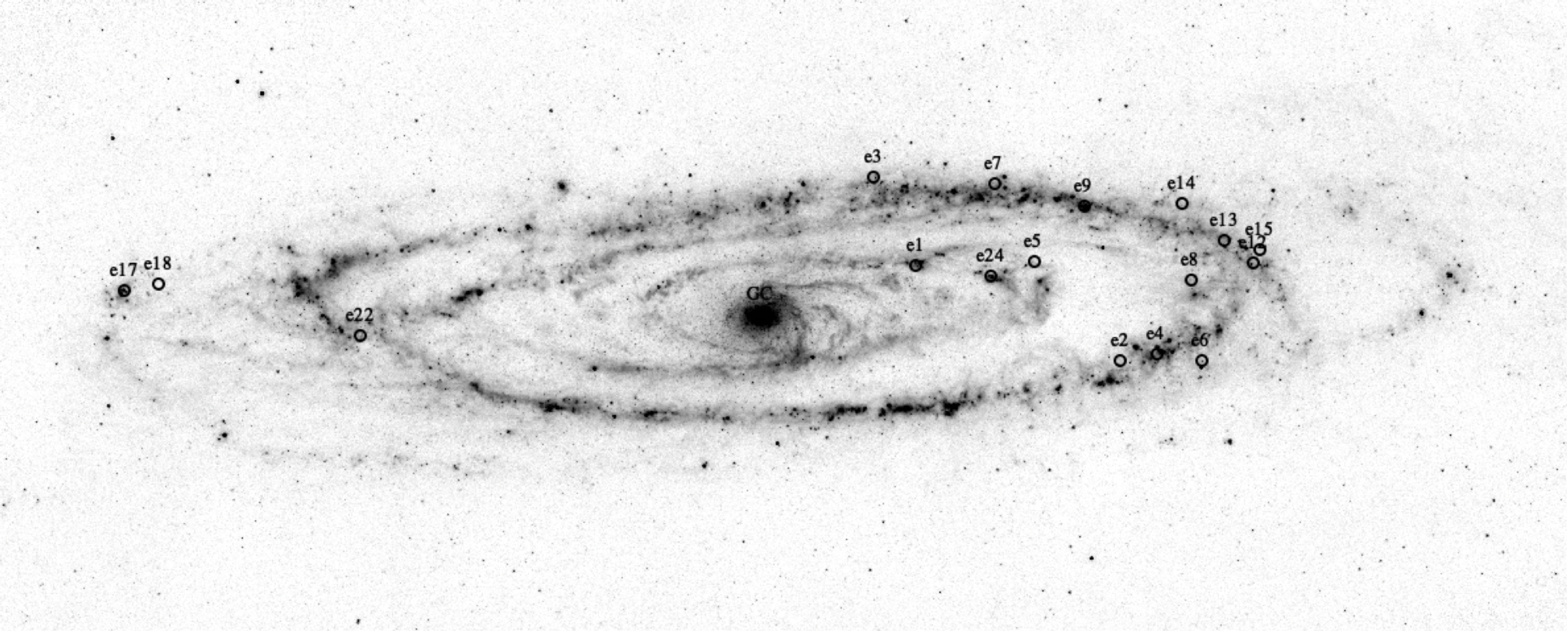}
\includegraphics[width=3.2in]{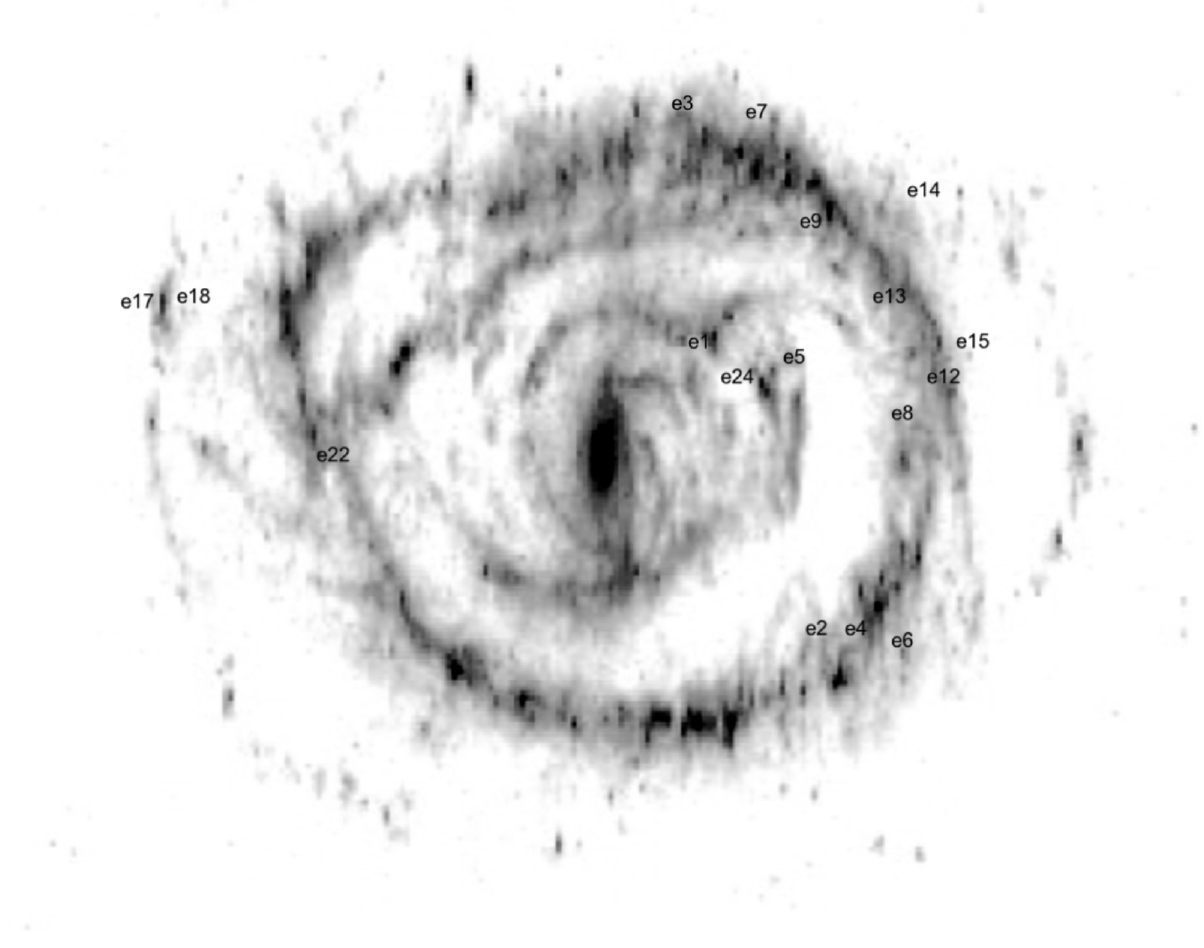}
\caption{The star locations are shown on the {\it Spitzer}/MIPS 24~\micron\ image of M31 \citep{2006ApJ...638L..87G} as it appears on the sky (top) and deprojected assuming an inclination of 78\arcdeg\ (bottom).}
\label{fig_region1}
\end{figure}

The target stars for the new observations, listed in Table~\ref{tb_targets} and shown in Fig.~\ref{fig_region1}, were chosen primarily from a new large list of M31 OB supergiants with accurate spectral types \citep{2016AJ....152...62M}.
In order to get good matches to the stellar model atmospheres used for comparison, the range of spectral types for the target sample was limited to those earlier than B3. For later type stars, the slope of the UV continuum changes more rapidly with spectral type making the pair method more difficult.
The \ebv\ value for each star was required to be greater than 0.25 so that the extinction internal to M31 is significantly larger than any MW foreground extinction.
The four sightlines studied in \citet{2015ApJ...815...14C}, e17, e18, e22, and e24, have been re-analyzed as part of this study.

Table~\ref{tb_targets} also gives the deprojected M31 galactocentric distance of each of the target stars assuming the distance to M31 is 785 kpc.
The coordinates for the center of M31 are assumed to be $\alpha$ = 00:42:44.32, $\delta$ = 41:16:08.5, and the inclination and position angle of the galaxy are 78$\fdg$1 and 37$\fdg$2 \citep{2010A&A...511A..89C}.
As shown in Figure~\ref{fig_region1}, the target stars have been selected to map out a variety of M31 galactocentric distances as well as to lie in the part of M31 with existing multi-band HST photometry \citep{2012ApJS..200...18D,2012AJ....144..142B}.

\subsection{Data}

Low resolution {\it HST}/STIS spectra in the G140L and G230L gratings (1150--3180~\AA\ at $R = 500-1000$) were obtained in GO program 14761 of 15 reddened OB stars that also lie in the footprint of the Panchromatic Hubble Andromeda Treasury (PHAT) survey in M31 \citep{2012ApJS..200...18D}. 
The PHAT survey covered an area of 0.5 deg$^2$ providing HST Wide Field Camera 3 (WFC3) and Advanced Camera for Surveys (ACS) photometry in the near-UV (F275W, F336W), optical (F475W, F814W), and near-infrared (F110W, F160W) bands \citep{2014ApJS..215....9W}.
The target stars lie in crowded fields in M31, so the observations were made with the 0.2$\arcsec$ $\times$ 0.2$\arcsec$ slit. Each target had a single 2-orbit visit with no repeat visits, one orbit for each grating. 
The observations for two of the stars in the GO-14761 program failed, therefore they were excluded from analysis (formally e10 and e11).

Because of the faintness of the M31 stars with net signals comparable to or occasionally below the sky background level, special data reduction techniques are required for the STIS observations.
First, the standard G230L dark signal subtraction is not precise enough, so an adjustment is constructed from the 2048x2048 pixel raw science image itself.
The science image is binned into eight bins in the row direction (X) by eight bins in the column direction (Y) with 32 blocks above the spectral trace and 32 below.
These 64 bins avoid the image edges by 104 pixels and avoid the spectral trace by 100 pixels.
For G230L, the average signal in each set of eight blocks in the X direction is fit with a fourth-order polynomial and that fit is evaluated for each of the 2048 pixels at the center of each row of eight blocks.
This set of 2048x8 fitted vectors is then fit at the eight Y positions for each of the 2048 columns by a third-order polynomial.
The evaluation of the 2048 Y fits at each of the 2048 Y pixels produces the new dark file for subtraction from the raw image.
A similar process produces a
revised G140L dark file, except for fitting in the X direction by linear interpolation.

In addition to the revised dark image, the standard sky background position moves from the standard distance (BDIST) of 300 rows from the spectral trace to BDIST=90 rows given the faintness of the spectra.
The absolute flux calibration changes slightly, because the closer BDIST includes a tiny fraction of the wings of the spectral line spread profile.

Finally, the spectral extraction uses an extraction width (GWIDTH) of three pixels in Y to avoid the extra noise in the standard GWIDTH=11 height.
To maintain an estimate of the absolute flux, the 3-pixel extractions are normalized to the standard 11 pixel spectra using wavelength regions of maximum signal at 1250--1500~\AA\ for
G140L and 2300--3000~\AA\ for G230L.
For example, for the j004546.81+415431.7 G140L spectrum this normalization increases the GWIDTH=3 spectrum a few percent to 23\% depending on the wavelength.
The 0.2 $\times$ 0.2$\arcsec$ aperture correction to get the absolute flux for the spectral SEDs is uncertain by 5\% RMS \citep{1998stis.rept...20B}.

The geocoronal emission from Ly$\alpha$ at 1215.67~\AA\ and from the much weaker \ion{O}{1} at 1302 -- 1306~\AA\ are imaged on top of the stellar signal with a width of $\sim$8 pixels that corresponds to the 0.2 $\times$ 0.2$\arcsec$ entrance slit size.
These contaminated regions of the G140L spectrum must be ignored in fitting models to the observations.

\begin{deluxetable*}{lccccccc}
\label{tb_phat}
\tablecaption{Photometry}
\tablehead{\colhead{ID} & \colhead{WFC3/F275W} & \colhead{WFC3/F336W} & \colhead{ACS/F475W}	& \colhead{ACS/F814W} & \colhead{WFC3/F110W} & \colhead{WFC3/F160W} }
\startdata
e1	&	18.032 $\pm$ 0.009	&	18.088 $\pm$ 0.006	&	19.153 $\pm$ 0.002	&	18.875 $\pm$ 0.001	&	18.957 $\pm$ 0.002	&	18.914 $\pm$ 0.003	\\
e2	&	17.643 $\pm$ 0.006	&	17.782 $\pm$ 0.01	&	19.018 $\pm$ 0.001	&	18.771 $\pm$ 0.001	&	18.916 $\pm$ 0.004	&	18.822 $\pm$ 0.002	\\
e3	&	17.684 $\pm$ 0.01	&	17.686 $\pm$ 0.007	&	18.801 $\pm$ 0.001	&	18.361 $\pm$ 0.001	&	18.307 $\pm$ 0.001	&	18.183 $\pm$ 0.002	\\
e4 &	18.737 $\pm$ 0.036	&	18.590 $\pm$ 0.026	&	19.355 $\pm$ 0.001	&	18.737 $\pm$ 0.001	&	18.602 $\pm$ 0.002	&	18.468 $\pm$ 0.002	\\
e5&	17.255 $\pm$ 0.01	&	16.932 $\pm$ 0.004	&	17.610 $\pm$ 0.017	&	16.996 $\pm$ 0.009	&	16.992 $\pm$ 0.002	&	16.864 $\pm$ 0.002	\\
e6&	16.367 $\pm$ 0.005	&	16.438 $\pm$ 0.005	&	17.555 $\pm$ 0.015	&	16.555 $\pm$ 0.008	&	15.578 $\pm$ 0.001	&	14.786 $\pm$ 0.001	\\
e7&	18.277 $\pm$ 0.013	&	18.139 $\pm$ 0.005	&	18.880 $\pm$ 0.001	&	18.009 $\pm$ 0.002	&	17.848 $\pm$ 0.001	&	17.625 $\pm$ 0.001	\\
e8&	18.167 $\pm$ 0.012	&	17.853 $\pm$ 0.009	&	18.534 $\pm$ 0.001	&	17.872 $\pm$ 0.001	&	17.814 $\pm$ 0.001	&	17.667 $\pm$ 0.001	\\
e9&	18.539 $\pm$ 0.015	&	18.681 $\pm$ 0.007	&	19.862 $\pm$ 0.002	&	19.504 $\pm$ 0.002	&	19.493 $\pm$ 0.011	&	19.341 $\pm$ 0.009	\\
e12&	19.249 $\pm$ 0.014	&	19.033 $\pm$ 0.009	&	19.869 $\pm$ 0.001	&	19.153 $\pm$ 0.001	&	19.036 $\pm$ 0.002	&	18.882 $\pm$ 0.002	\\
e13&	19.225 $\pm$ 0.012	&	18.952 $\pm$ 0.007	&	19.626 $\pm$ 0.002	&	18.904 $\pm$ 0.002	&	18.809 $\pm$ 0.003	&	18.661 $\pm$ 0.002	\\
e14&	19.230 $\pm$ 0.014	&	18.816 $\pm$ 0.011	&	19.385 $\pm$ 0.001	&	18.544 $\pm$ 0.001	&	18.425 $\pm$ 0.003	&	18.226 $\pm$ 0.002	\\
e15&	18.065 $\pm$ 0.016	&	17.994 $\pm$ 0.009	&	18.959 $\pm$ 0.001	&	18.466 $\pm$ 0.001	&	18.397 $\pm$ 0.001	&	18.282 $\pm$ 0.002	\\
e22 & \nodata & 17.686 $\pm$ 0.006 & 18.871 $\pm$ 0.001 & 18.415 $\pm$ 0.001 & \nodata & \nodata \\
e24 & 17.157 $\pm$ 0.005 & 16.842 $\pm$ 0.002 & \nodata & 16.826 $\pm$ 0.006 & 16.757 $\pm$ 0.001 & 16.563 $\pm$ 0.001\\ \hline \hline
\colhead{ID} & \colhead{WFPC2/F170W} & \colhead{WFPC2/F255W} & \colhead{WFPC2/F336W}	& \colhead{WFPC2/439W} & \colhead{WFPC2/F555W} & \colhead{WFPC2/F814W} \\ \hline
e17 & $17.30 \pm 0.02$ & $17.42 \pm 0.05$ & $17.32 \pm 0.01$ & $18.62 \pm 0.01$ & $18.42 \pm 0.01$ & $18.09 \pm 0.01$ \\
e22 & $17.74 \pm 0.04$ & $17.65 \pm 0.07$ & $17.67 \pm 0.01$ & $18.93 \pm 0.01$ & $18.73 \pm 0.01$ & $18.39 \pm 0.01$ \\
e24 & $17.99 \pm 0.06$ & $17.34 \pm 0.06$ & $16.75 \pm 0.01$ & $17.74 \pm 0.01$ & $17.36 \pm 0.01$ & $16.81 \pm 0.01$ \\ \hline \hline
\colhead{ID} & \colhead{V} & \colhead{B-V} & \colhead{U-B} & \colhead{V-R} & \colhead{R-I} & \\ \hline
e18 & $18.986 \pm 0.003$ & $0.095 \pm 0.003$ & $-0.900 \pm 0.003$ & $0.107 \pm 0.004$ & $0.068 \pm 0.003$ \\
\enddata
\end{deluxetable*}

The photometry for all the stars is listed in Table~\ref{tb_phat}. 
The majority of stars have PHAT photometry.
For the stars from \citet{2015ApJ...815...14C}, e17, e22, and e24 have HST/WFPC2 photometry \citep{2012AJ....144..142B}, e22 has HST photometry from the PHAST survey \citep{2025ApJ...979...35C}, and e18 only has ground-based photometry \citep{2006AJ....131.2478M}.

\begin{figure}[tbp]
\epsscale{1.2}
\plotone{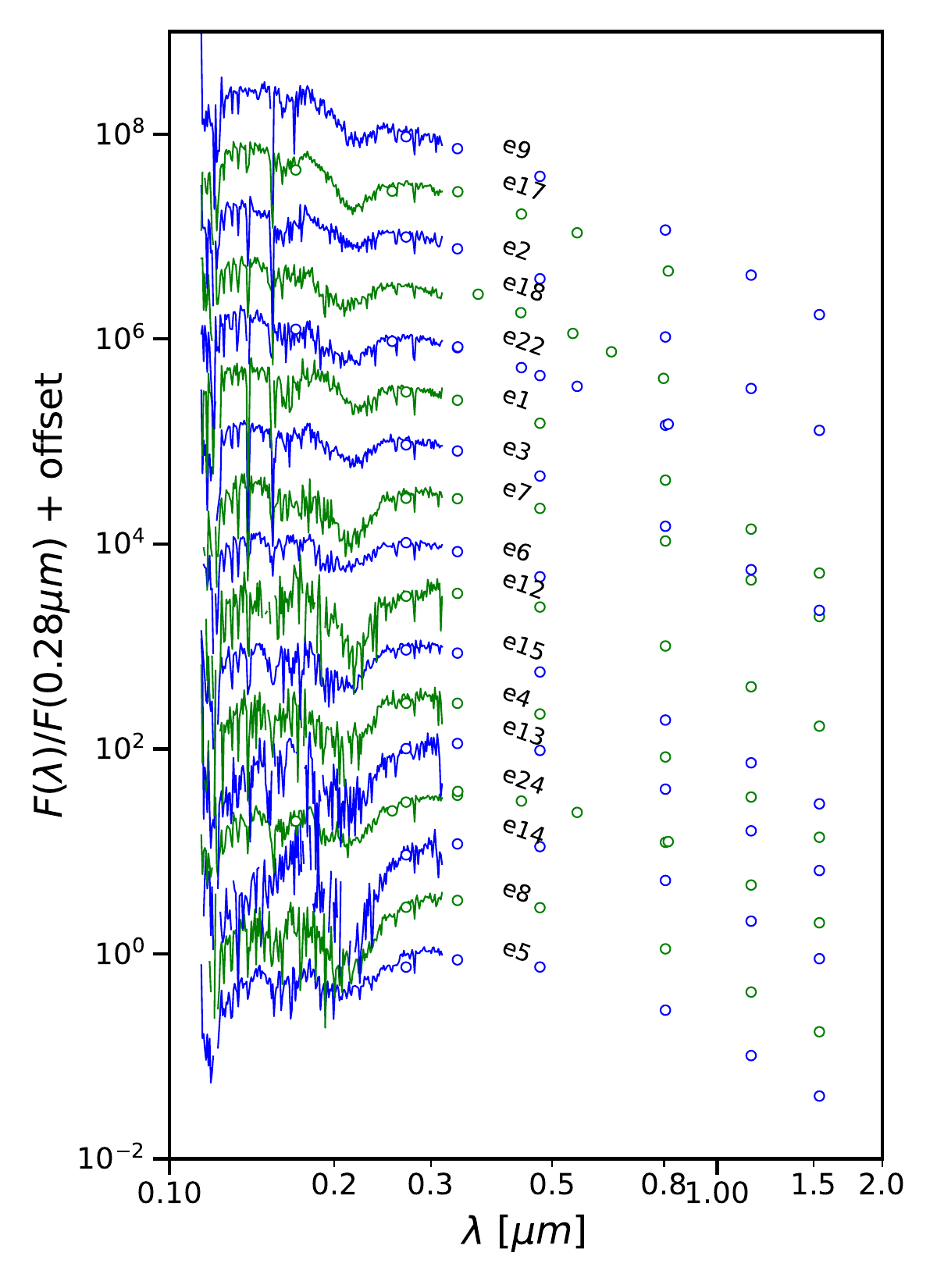}
\caption{The STIS spectra and photometry for the target stars are shown.
The spectra have been normalized to the flux at 2800~\AA, offset, and sorted by the UV spectral slope.
Regions of anomalously low flux have not been plotted.
\label{fig_spectra}}
\end{figure}

The STIS spectra and photometry are plotted in Figure~\ref{fig_spectra}. 

\subsection{Extinction Curves}

\begin{figure*}[tbp]
\epsscale{1.2}
\plotone{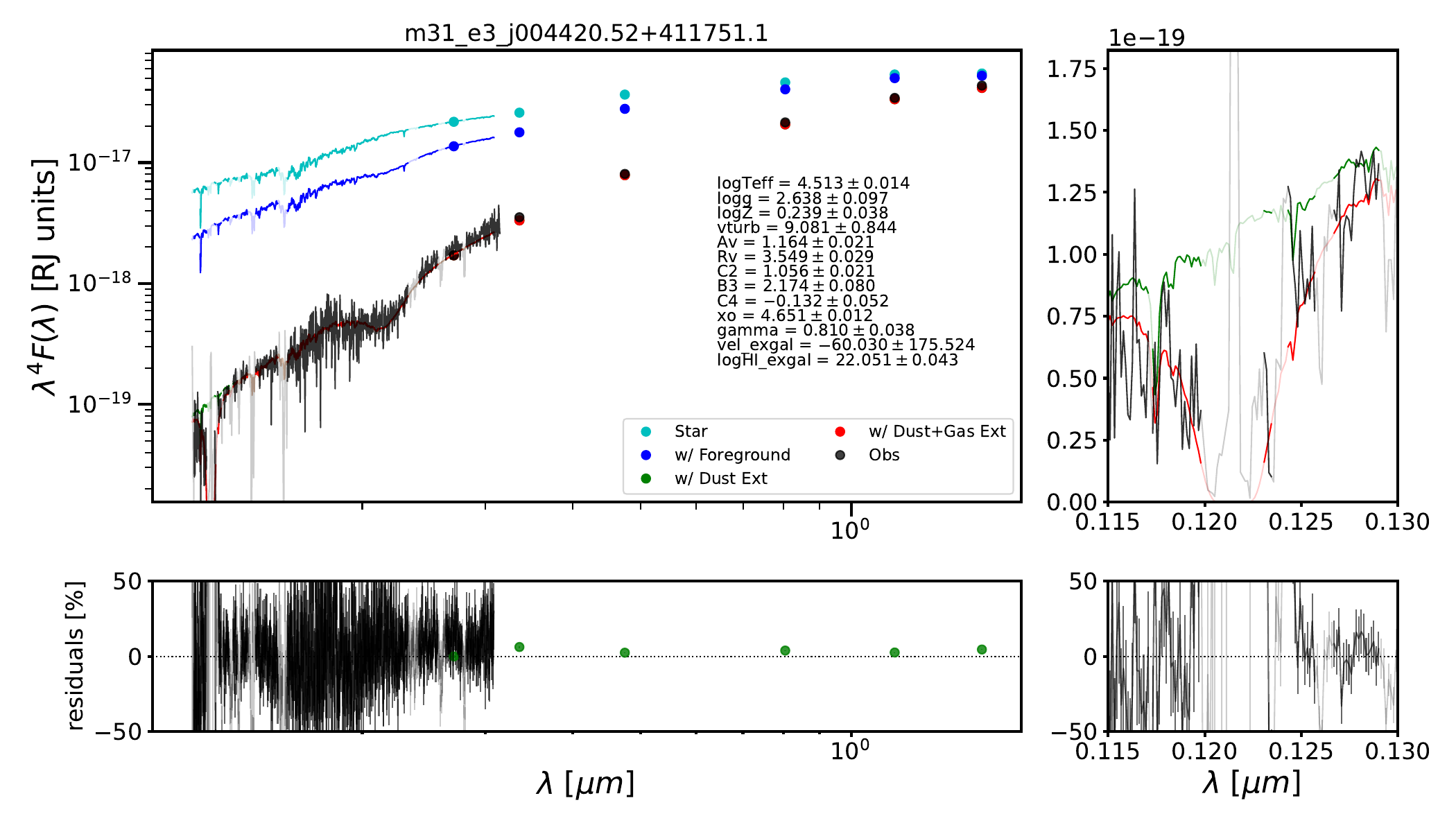}
\label{fig_exp_fit}
\caption{The observed data for J004420.52+411751.1 (e3) are shown in black in the top panels.
In the left top panel, the full wavelength range is shown with the unextinguished model at the top (cyan), the MW foreground extinguished model in the middle (blue), and the MW foreground and M31 internal dust extinguished model overplotted on the observations (red).
The right top panel gives the region around Ly$\alpha$ with the model with dust extinction but not gas absorption in green and the full model including the gas absorption in red.
The bottom panels give the residuals between the observations and the full model.
The numerical values of the fit parameters are given in the upper left panel.
\label{fig_exp_fit}}
\end{figure*}
 
We measured stellar, extinction, and \ion{H}{1} column parameters towards each star by forward modeling the spectra and photometry using the `measure\_extinction' python package \citep{measureextinction}.
This is an updated version of the fitting done for the previous M31 extinction analysis \citep{2015ApJ...815...14C}.
The stellar atmosphere models used were those from the non-LTE Tlusty model \citep{Lanz03, 2025AJ....169..178H} and the unreddened simulated STIS spectroscopy was convolved with the appropriate wavelength-dependent line spread functions.
Tlusty models with metallicites ranging from 1/5 to 2 $\times$ Solar were used to get the best fit to the observed spectra giving a rough idea of the metallicity of the individual stars. 
The stellar models were extinguished using the FM90 functional form \citep{1990ApJS...72..163F} for $\lambda <$ 2700~\AA, the G23 $\rv$ dependent model \citep{2023ApJ...950...86G} for $\lambda >$ 4000~\AA, and a cubic spline interpolation to connect the two \citep{2019ApJ...886..108F}.
We use a variant of the FM90 fitting where the 2175 \AA\ bump amplitude is given by B$_3$ = C$_3$/$\gamma^2$ as B$_3$ directly measures the bump amplitude unlike C$_3$ \citep{2024ApJ...970...51G}.
The `dust\_extinction' package \citep{2024JOSS....9.7023G} provided the FM90 model and the G23 model \citep{2023ApJ...950...86G} which is based on four literature studies \citep{Gordon:2009fk, 2019ApJ...886..108F, 2021ApJ...916...33G, 2022ApJ...930...15D}.
For each reddened model, the photometry was calculated by integrating over the appropriate band response functions.
All the model parameters are given in Table~\ref{tab_fit_params} along with the minimum and maximum values of the square priors.
For some of the parameters, Gaussian priors were also used and the center and width is given in the same table.

\begin{deluxetable}{clcccc}
\tablecaption{Model Parameters}
\tablehead{  & & & & \colhead{Gaussian} \\
 \colhead{Parameter} & \colhead{Description} &
 \colhead{Units} &
    \colhead{Min} &
    \colhead{Max} &
    \colhead{(center, $\sigma$})}
\startdata  
\multicolumn{5}{c}{M31 Components} \\ \hline
$\log(T_\mathrm{eff})$ & effective temperature & K & 4.18 & 4.74 & (x\tablenotemark{a}, 0.025)\\
$\log(g)$  & surface gravity & cm s$^{-1}$ & 1.75 & 4.75 & (y\tablenotemark{a}, 0.1) \\
$\log(Z)$  & metallicity & & -0.7 & 0.3 & (0, 0.2) \\
$v_\mathrm{turb}$ & turbulent velocity & km s$^{-1}$ & 2 & 10 & \nodata \\
$A(V)$  & V band extinction & mag & 0.0 & 100.0 & \nodata \\
$R(V)$ & $A(V)/E(B-V)$ & & 2.3 & 5.6 & (3.0, 0.4) \\
$C_2$ & UV slope & & -0.1 & 5.0 & (0.73, 0.25) \\
$B_3$ & 2175~\AA\ bump height & & -1.0 & 8.0 & (3.6, 0.6) \\
$C_4$ & FUV curvature & & -0.5 & 1.5 & (0.4, 0.2) \\
$x_o$ & 2175~\AA\ bump centroid & $\micron^{-1}$ & 4.5 & 4.9 & (4.59, 0.2) \\
$\gamma$ & 2175~\AA\ bump width & $\micron^{-1}$ & 0.4 & 1.7 & (0.89, 0.08) \\ 
$\log(HI)$ & M31 H I column & atoms cm$^{-2}$ & 16.0 & 24.0 & \nodata \\ 
$v(M31)$  & velocity & km s$^{-1}$ & \multicolumn{3}{c}{fixed, $-300$} \\ \hline
\multicolumn{5}{c}{MW Components} \\ \hline
$\log_\mathrm{MW}(H I)$ & MW H I column & atoms cm$^{-2}$ & \multicolumn{3}{c}{fixed, Table~\ref{tab_ext_col_param}} \\
$E(B-V)_{MW}$ & MW dust column & mag & \multicolumn{3}{c}{fixed, Table~\ref{tab_ext_col_param}} \\
$R(V)_{MW}$ & MW $A(V)/E(B-V)$ & & \multicolumn{3}{c}{fixed, 3.1} \\
$v(MW)$  & velocity & km s$^{-1}$ & \multicolumn{3}{c}{fixed, 0} 
\enddata
\tablenotemark{a}{Gaussian centers x and y for $\log(T_\mathrm{eff})$ and $\log(g)$, respectively, were set by the spectral types given in Table~\ref{tb_targets}.}
\label{tab_fit_params}
\end{deluxetable}

As part of this modeling, the expected MW foreground dust extinction and Ly$\alpha$ \ion{H}{1} absorption are included through a model based on the radio measured \ion{H}{1} gas column, the corresponding estimated dust column, and assuming a wavelength dependence given by the G23 $R(V)=3.1$ model \citep{2023ApJ...950...86G}.
The dust portion of this model has been shown to be a good fit for foreground dust extinction towards the SMC \citep{2024ApJ...970...51G}.
Estimates of the MW foreground dust column toward M31 give an average $E(B - V) \sim 0.06$ mag \citep{1998ApJ...500..525S,2011ApJ...737..103S,2011ApJ...726...39C,2012AJ....144..142B}.
More accurate estimates of the dust column for each sightline can be determined using radio \ion{H}{1} measurements integrated over the range of MW velocities and the measured high-latitude MW \nhiebv\ ratio.
The $N(HI)_{MW}$ values for each star were measured by integrating H I spectra from the HI4Pi survey at velocities between -80 and 70~km~$s^{-1}$ (local MW HI only) \citep{2016A&A...594A.116H}.
Using the high-latitude MW measured \nhiebv\ value of $8.3 \times 10^{21}$~H~cm$^{-2}$~mag$^{-1}$ \citep{Liszt14} these were converted to $E(B-V)$ values.
Assuming the MW foreground has $R(V)=3.1$ and the corresponding G23 wavelength dependence \citep{2023ApJ...950...86G}, provides the model for the MW foreground extinction at UV through near-IR wavelengths.
The corresponding foreground $A(V)_\mathrm{MW}$ values range from 0.19 to 0.25~mag.
The MW \ion{H}{1} Ly$\alpha$ absorption is modeled with a Voigt profile with the radio measured \ion{H}{1} column and a velocity set to 0~km~s$^{-1}$.
As this is explicitly part of the forward model, the dust extinction and \ion{H}{1} parameters derived only measure the M31 dust and gas properties.

The fitting was done using a Bayesian framework.
By default, most fit parameters had square priors imposed to keep the parameters inside the defined model grid (stellar parameters) and positive (many of the gas and dust parameters). 
In addition, Gaussian priors were applied for the $\log(T\mathrm{eff})$ and $\log(g)$ parameters with the centers based on the literature stellar spectral types (see Table~\ref{tb_targets}) and widths of 0.025 and 0.1, respectively.
It was necessary to impose an additional square prior on both parameters with min and max values based on mean $\pm 3\sigma$ to avoid results very far from the known spectral types.
For the \rv\ and FM90 parameters, Gaussian priors based on MW distributions of these parameters were used \citep{2004ApJ...616..912V, 2007ApJ...663..320F}.
The posterior PDFs of the parameters were determined using Markov Chain Monte Carlo (MCMC) sampling, specifically with the `emcee' package \citep{2013PASP..125..306F, 2019JOSS....4.1864F} with enough steps (100,000) that the walkers had relaxed.

During the fitting, we found for e6 that the F814W, F110W, and F160W photometry showed an excess above any reasonable model.
The origin of this excess is not known and could be a companion star or a disk.
We excluded these bands from the fitting for this star.

\begin{deluxetable}{lcccc}
\tablecaption{Stellar Parameters \label{tab_ext_stell_params}}
\tablehead{\colhead{Name} & \colhead{$\log(T_\mathrm{eff})$} & \colhead{$\log(g)$} & \colhead{$\log(Z)$} & \colhead{$v_\mathrm{turb}$}}
\startdata
e1 & $4.459 \pm 0.012$ & $3.506 \pm 0.056$ & $0.123 \pm 0.106$ & $9.08 \pm 1.04$ \\
e2 & $4.491 \pm 0.041$ & $2.440 \pm 0.128$ & $-0.095 \pm 0.162$ & $9.24 \pm 0.67$ \\
e3 & $4.513 \pm 0.014$ & $2.638 \pm 0.097$ & $0.239 \pm 0.038$ & $9.08 \pm 0.84$ \\
e4 & $4.327 \pm 0.013$ & $2.324 \pm 0.090$ & $0.198 \pm 0.054$ & $8.97 \pm 0.87$ \\
e5 & $4.185 \pm 0.012$ & $2.441 \pm 0.075$ & $0.248 \pm 0.046$ & $9.56 \pm 0.65$ \\
e6 & $4.230 \pm 0.000$ & $2.001 \pm 0.001$ & $-0.000 \pm 0.000$ & $5.00 \pm 0.00$ \\
e7 & $4.395 \pm 0.005$ & $2.529 \pm 0.107$ & $-0.000 \pm 0.003$ & $3.77 \pm 0.44$ \\
e8 & $4.275 \pm 0.004$ & $2.015 \pm 0.014$ & $0.184 \pm 0.047$ & $9.73 \pm 0.41$ \\
e9 & $4.512 \pm 0.015$ & $3.305 \pm 0.094$ & $-0.040 \pm 0.135$ & $8.75 \pm 0.92$ \\
e12 & $4.350 \pm 0.007$ & $3.276 \pm 0.033$ & $0.184 \pm 0.142$ & $8.05 \pm 0.41$ \\
e13 & $4.312 \pm 0.012$ & $2.320 \pm 0.094$ & $0.038 \pm 0.151$ & $8.69 \pm 2.35$ \\
e14 & $4.345 \pm 0.011$ & $2.376 \pm 0.100$ & $0.181 \pm 0.046$ & $9.05 \pm 0.82$ \\
e15 & $4.316 \pm 0.002$ & $2.272 \pm 0.065$ & $0.151 \pm 0.154$ & $7.55 \pm 0.03$ \\
e17 & $4.486 \pm 0.016$ & $2.457 \pm 0.083$ & $-0.110 \pm 0.155$ & $8.93 \pm 0.83$ \\
e18 & $4.362 \pm 0.003$ & $2.498 \pm 0.014$ & $0.000 \pm 0.002$ & $5.00 \pm 0.02$ \\
e22 & $4.341 \pm 0.008$ & $2.011 \pm 0.012$ & $-0.146 \pm 0.162$ & $7.54 \pm 0.04$ \\
e24 & $4.286 \pm 0.009$ & $2.506 \pm 0.091$ & $0.020 \pm 0.103$ & $8.79 \pm 0.80$
\enddata
\end{deluxetable}

\begin{deluxetable}{lccccccc}
\tablecaption{Column Parameters \label{tab_ext_col_param}}
\tablehead{\colhead{Name} & \colhead{$A(V)$} & \colhead{$R(V)$} & \colhead{$log[N(H I)]$} & \colhead{$A(V)_\mathrm{MW}$} & \colhead{$log[N(H I)]_\mathrm{MW}$}}
\startdata
e1 & $0.75 \pm 0.01$ & $2.33 \pm 0.02$ & $21.94 \pm 0.04$ & $0.24$ & $20.81$ \\
e2 & $0.58 \pm 0.03$ & $2.34 \pm 0.04$ & $21.90 \pm 0.04$ & $0.25$ & $20.82$ \\
e3 & $1.16 \pm 0.02$ & $3.55 \pm 0.03$ & $22.05 \pm 0.04$ & $0.25$ & $20.82$ \\
e4 & $1.23 \pm 0.02$ & $3.05 \pm 0.03$ & \nodata & $0.25$ & $20.82$ \\
e5 & $1.05 \pm 0.02$ & $3.30 \pm 0.07$ & $21.52 \pm 0.20$ & $0.25$ & $20.82$ \\
e6 & $0.36 \pm 0.04$ & $2.70 \pm 0.29$ & $21.68 \pm 0.05$ & $0.25$ & $20.82$ \\
e7 & $1.72 \pm 0.03$ & $3.36 \pm 0.02$ & $22.03 \pm 0.10$ & $0.25$ & $20.83$ \\
e8 & $1.01 \pm 0.00$ & $2.48 \pm 0.01$ & \nodata & $0.24$ & $20.81$ \\
e9 & $1.21 \pm 0.02$ & $3.98 \pm 0.08$ & $21.68 \pm 0.08$ & $0.25$ & $20.82$ \\
e12 & $1.40 \pm 0.01$ & $2.77 \pm 0.02$ & \nodata & $0.24$ & $20.81$ \\
e13 & $1.19 \pm 0.03$ & $2.48 \pm 0.03$ & \nodata & $0.24$ & $20.80$ \\
e14 & $1.38 \pm 0.02$ & $2.47 \pm 0.02$ & \nodata & $0.24$ & $20.81$ \\
e15 & $0.98 \pm 0.01$ & $2.98 \pm 0.03$ & $21.81 \pm 0.11$ & $0.24$ & $20.81$ \\
e17 & $1.43 \pm 0.07$ & $4.87 \pm 0.26$ & $21.97 \pm 0.03$ & $0.19$ & $20.70$ \\
e18 & $0.85 \pm 0.04$ & $4.11 \pm 0.23$ & $21.97 \pm 0.04$ & $0.19$ & $20.70$ \\
e22 & $1.43 \pm 0.00$ & $5.59 \pm 0.01$ & $22.03 \pm 0.04$ & $0.19$ & $20.71$ \\
e24 & $1.50 \pm 0.02$ & $3.82 \pm 0.04$ & $21.87 \pm 0.07$ & $0.25$ & $20.82$
\enddata
\end{deluxetable}

\begin{deluxetable}{lcccccc}
\tablecaption{FM90 Parameters \label{tab_ext_fm90_params}}
\tablehead{\colhead{Name} & \colhead{$C_2$} & \colhead{$B_3$} & \colhead{$C_4$} & \colhead{$x_o$ [\micron$^{-1}$]} & \colhead{$\gamma$ [\micron$^{-1}$]}}
\startdata
e1 & $1.26 \pm 0.04$ & $2.06 \pm 0.08$ & $-0.24 \pm 0.07$ & $4.566 \pm 0.015$ & $0.91 \pm 0.06$ \\
e2 & $1.03 \pm 0.07$ & $1.94 \pm 0.10$ & $-0.07 \pm 0.06$ & $4.629 \pm 0.023$ & $0.64 \pm 0.04$ \\
e3 & $1.06 \pm 0.02$ & $2.17 \pm 0.08$ & $-0.13 \pm 0.05$ & $4.651 \pm 0.012$ & $0.81 \pm 0.04$ \\
e4 & $1.09 \pm 0.05$ & $2.11 \pm 0.15$ & $0.10 \pm 0.10$ & $4.591 \pm 0.017$ & $0.77 \pm 0.06$ \\
e5 & $1.36 \pm 0.03$ & $1.64 \pm 0.07$ & $0.04 \pm 0.07$ & $4.608 \pm 0.016$ & $0.92 \pm 0.04$ \\
e6 & $1.89 \pm 0.11$ & $4.24 \pm 0.25$ & $-0.16 \pm 0.08$ & $4.702 \pm 0.011$ & $0.88 \pm 0.03$ \\
e7 & $0.76 \pm 0.02$ & $2.31 \pm 0.10$ & $-0.09 \pm 0.06$ & $4.624 \pm 0.013$ & $0.81 \pm 0.04$ \\
e8 & $1.35 \pm 0.03$ & $2.99 \pm 0.15$ & $-0.27 \pm 0.11$ & $4.568 \pm 0.019$ & $0.88 \pm 0.04$ \\
e9 & $0.53 \pm 0.04$ & $1.93 \pm 0.11$ & $0.39 \pm 0.07$ & $4.590 \pm 0.016$ & $0.86 \pm 0.06$ \\
e12 & $0.86 \pm 0.04$ & $1.84 \pm 0.25$ & $0.14 \pm 0.09$ & $4.581 \pm 0.019$ & $0.85 \pm 0.07$ \\
e13 & $1.11 \pm 0.05$ & $2.08 \pm 0.21$ & $0.28 \pm 0.13$ & $4.582 \pm 0.018$ & $1.03 \pm 0.07$ \\
e14 & $1.18 \pm 0.05$ & $2.64 \pm 0.31$ & $0.22 \pm 0.12$ & $4.591 \pm 0.021$ & $0.99 \pm 0.07$ \\
e15 & $1.17 \pm 0.03$ & $2.80 \pm 0.13$ & $-0.09 \pm 0.08$ & $4.635 \pm 0.014$ & $0.65 \pm 0.05$ \\
e17 & $0.69 \pm 0.04$ & $3.65 \pm 0.09$ & $0.27 \pm 0.06$ & $4.614 \pm 0.008$ & $0.92 \pm 0.03$ \\
e18 & $1.45 \pm 0.03$ & $4.07 \pm 0.14$ & $-0.04 \pm 0.08$ & $4.671 \pm 0.014$ & $1.24 \pm 0.05$ \\
e22 & $0.85 \pm 0.06$ & $2.96 \pm 0.12$ & $-0.09 \pm 0.07$ & $4.656 \pm 0.018$ & $0.71 \pm 0.04$ \\
e24 & $1.36 \pm 0.03$ & $2.34 \pm 0.06$ & $-0.20 \pm 0.06$ & $4.608 \pm 0.014$ & $1.16 \pm 0.04$ \\
Average & $1.21 \pm 0.02$ & $3.00 \pm 0.05$ & $0.13 \pm 0.03$ & $4.623 \pm 0.007$ & $1.04 \pm 0.04$
\enddata
\end{deluxetable}

The fits for the stellar parameters are given in Table~\ref{tab_ext_stell_params}, for the dust and gas column parameters in Table~\ref{tab_ext_col_param}, and for the FM90 UV extinction parameters in Table~\ref{tab_ext_fm90_params}.
For some of the sightlines, the data around Ly$\alpha$ were too noisy to measure the \ion{H}{1} column density and these are indicated by no data in the appropriate column in Table~\ref{tab_ext_col_param}.
Fig.~\ref{fig_exp_fit} shows the observations and model for e3 and illustrates the unextinguished, MW foreground extinguished, MW+M31 dust extinguished, and MW+M31 dust and gas extinguished models.

The extinction curves were calculated using the model based on the median (50\%) posterior PDF fit parameters stellar plus MW foreground extinction model.
Measuring the extinction curve this way is a variant of the standard pair method \citep{1983ApJ...266..662M} for measuring extinction curves where a reddened star is compared to a comparison star of the same spectral type allowing the effects of dust to be isolated.
In this case, a stellar atmosphere model combined with a MW foreground extinction model is used for the comparison instead of a measured unreddened star \citep{2005AJ....130.1127F,2019ApJ...886..108F}.
This method is sensitive to the absolute flux calibration of the observations, but has the advantage of not requiring observations of unreddened comparison stars.

\begin{figure}
\epsscale{1.2}
\plotone{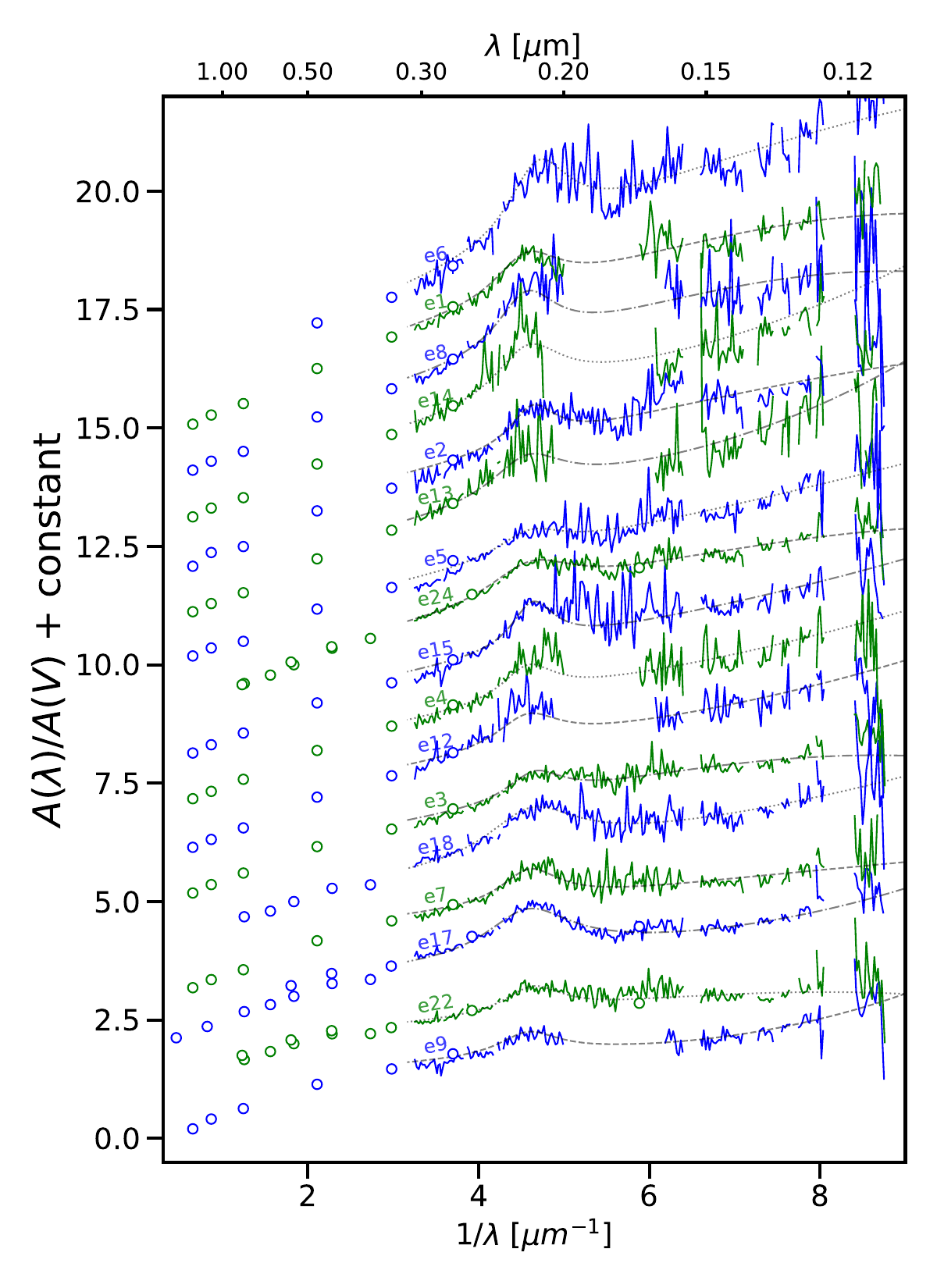}
\caption{The MW foreground-corrected M31 extinction curves are plotted sorted by UV slope in these units (i.e., $C_2 / R(V) + 1$).
The curves have been rebinned to a resolution of 200.
The FM90 fits are plotted as non-solid lines.
For clarity, the curves have been offset on the y-axis.
Regions of low S/N, near Ly$\alpha$, and around wind lines have been masked.
\label{fig_curves}}
\end{figure}

Most of the sightlines do not have V band photometry, so the measured extinction was calculated relative to the F475W band giving unnormalized curves in $E(\lambda - \mathrm{F475W})$ for sightlines.
These curves were converted to the more standard $E(\lambda - V)$ units by adding $E(\mathrm{F475W} - V)$ computed assuming G23 models and the \av\ and \rv\ best fit parameters.
Two sightlines had neither $V$ nor $F475W$ photometry and those were measured relative to $F439W$ and converted to relative to $V$ as above, with the appropriate adjustment for the different filter.
With all the extinction curves in $E(\lambda - V)$ it was straightforward to compute $A(\lambda)/A(V)$ using the \av\ best fit parameter.
These normalized curves are shown in Fig.~\ref{fig_curves} along with the FM90 models plotted as non-solid smooth lines.

\section{Discussion}

The aims of this study are to expand the sample of Local Group galaxies which have UV extinction observations to better understand how dust characteristics vary from galaxy to galaxy, and  across the face of a solar metallicity galaxy similar to the MW.
It is impossible to do this for the MW, itself, because we are sitting in the middle of the disk and can't see most of it in the UV.
In fact, most of the extinction sightlines studied in the MW, on which all \rv\ dependent relationships are based, are within roughly 1~kpc of the Sun \citep{2004ApJ...616..912V}.
The new observations presented here probe interstellar sightlines in M31 covering a wide area of that galaxy, having galactocentric distances of 5 to 16~kpc. 

\subsection{Comparison with Other Local Group Galaxies}

\begin{figure*}[tbp]
\epsscale{1.15}
\plotone{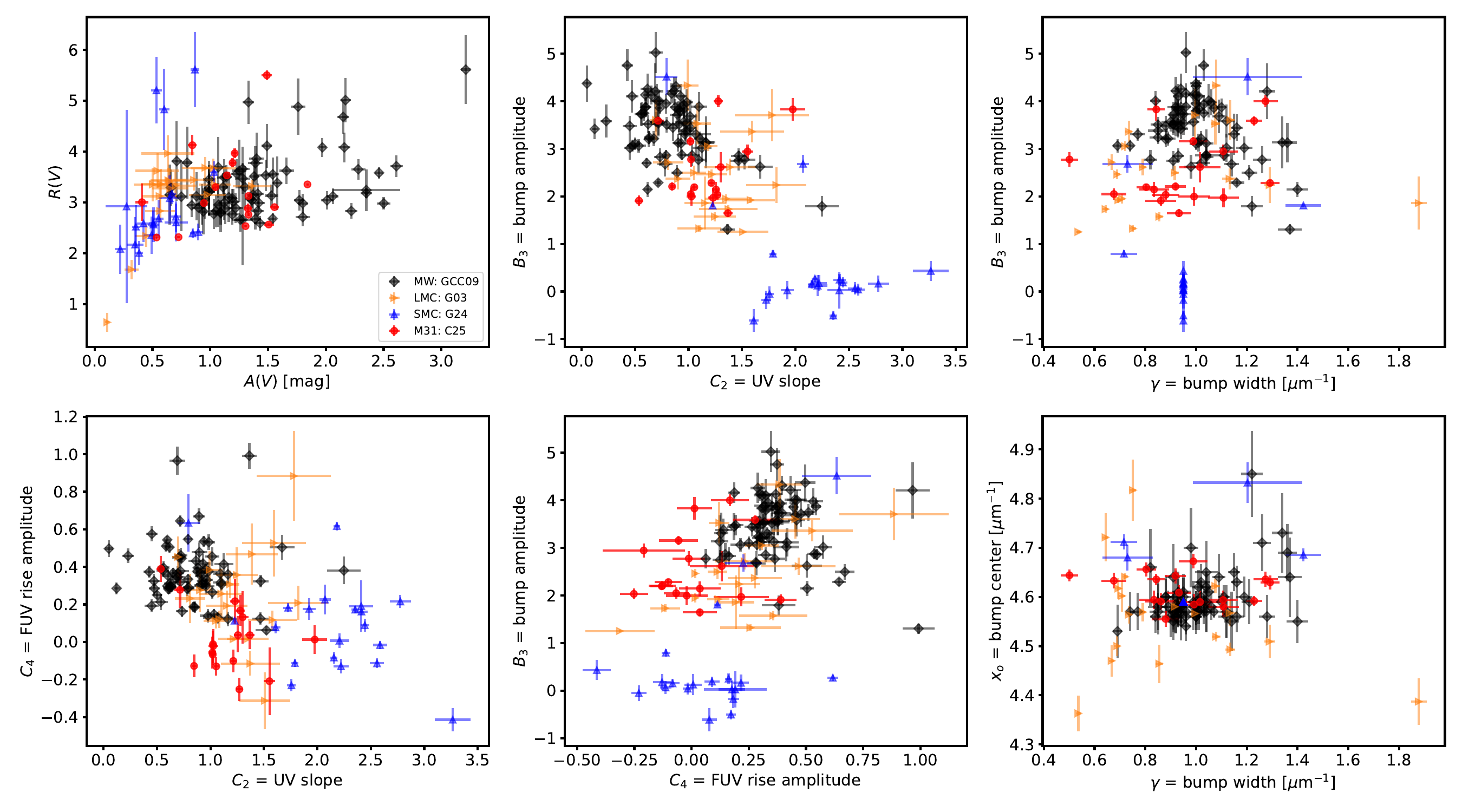}
\caption{\av\ versus \rv\ is shown in the upper left panel.
The other panels show different FM90 parameters versus each other.  
The $C_1$ versus $C_2$ correlation is not shown as our fitting technique does not include $C_1$, instead $C_1$ is related to $C_2$ using the known strong correlations between these two parameters.
\label{fig_fm90}}
\end{figure*}

\begin{figure*}[tbp]
\epsscale{1.15}
\plotone{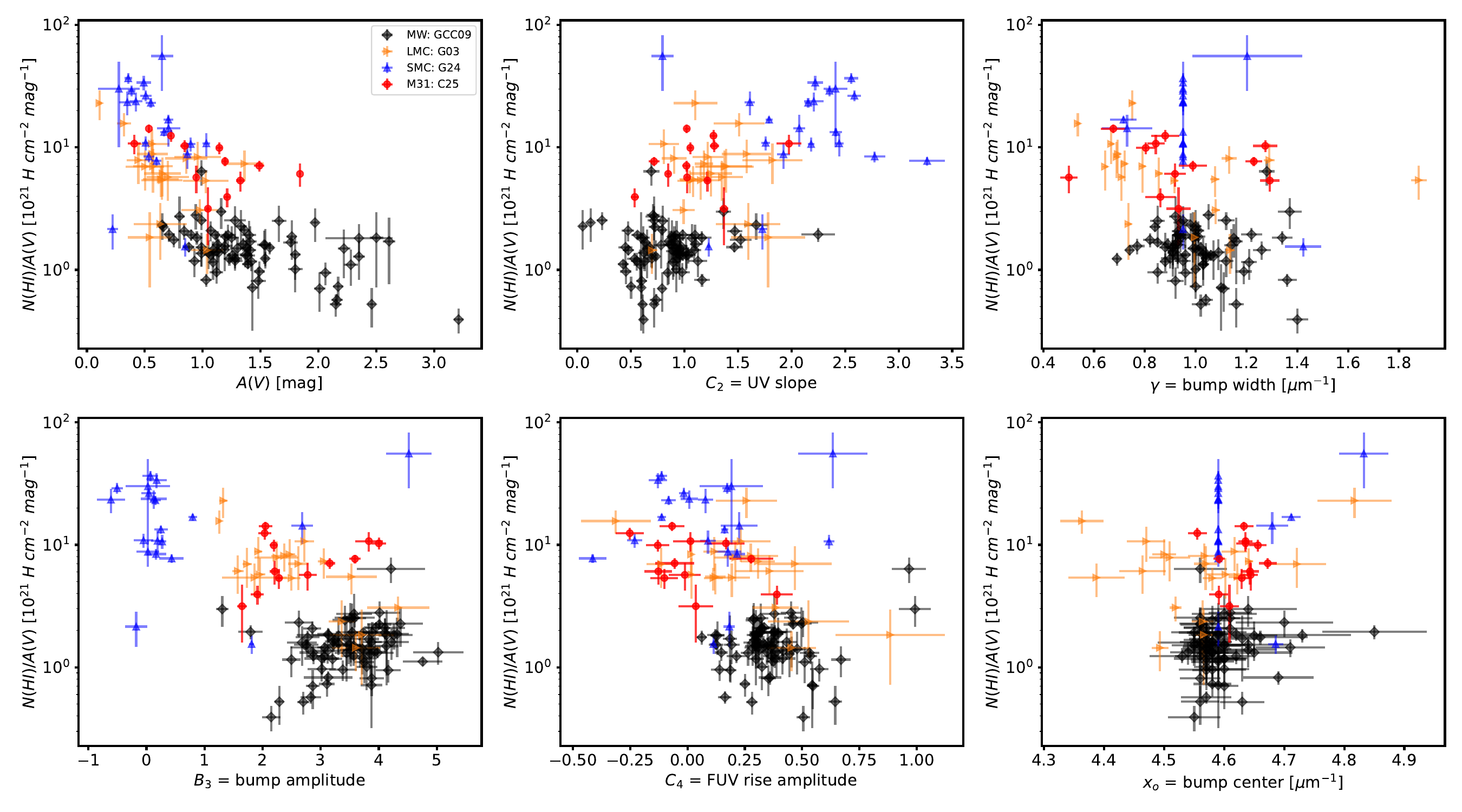}
\caption{The behavior of the FM90 extinction parameters and \av\ versus gas-to-dust ratio \nhiav\ are plotted.
The plot of \nhiav\ versus \av\ shows that generally the gas-to-dust ratio increases from the MW to the LMC to the SMC.
\label{fig_gdprops}}
\end{figure*}

As a first step, the observables for M31 dust, \av, \rv, \nhi, and the FM90 UV parameters can be compared to the same parameters measured along sightlines in the MW, LMC, and SMC.
Figures~\ref{fig_fm90} and \ref{fig_gdprops} show these various parameters plotted against each other for the MW \citep{Gordon:2009fk}, LMC \citep{2003ApJ...594..279G}, and SMC \citep{2024ApJ...970...51G}. 

In the upper left panel of Fig.~\ref{fig_fm90}, the behavior of \av\ versus \rv\ in M31 is seen to lie with the combined MW, LMC, and SMC points indicating that the M31 curves are probing similar dust columns and average grain sizes.
The rest of the panels in this figure show the correlations between the FM90 parameters.
Overall, the distribution of the M31 points lies within the distributions seen in the combined MW, LMC, and SMC points, with M31 overlapping most with the LMC points.
The correlations seen by \citet{2024ApJ...970...51G} for the MW, LMC, and SMC between $C_2$ (UV slope), $B_3$ (bump amplitude), and $C_4$ (FUV rise) are also seen in M31.
One difference is that the correlation in M31 between $C_4$ and $C_2$ is a bit steeper.
The upper right panel shows the correlation between $\gamma$ (the bump width) and $B_3$ that overall has a wedge shape with weaker bumps having a larger $\gamma$ variation.
The M31 points mainly populate the half of the wedge corresponding to narrower, lower $\gamma$ bumps.
The lower right panel shows the $\gamma$ versus $x_o$ (bump center) that does not show any significant correlation in M31 or any of the other galaxies.
In particular, the $C_4$ and $B_3$ values for M31 are very similar to those for the LMC2-30Dor sightlines.

In Figure~\ref{fig_gdprops}, the behavior of the \av\ and the FM90 extinction parameters versus the gas-to-dust ratio, \nhiav, is plotted.
The upper left panel shows that the \nhiav\ ratios for M31 are very similar to those seen in the LMC, and are overall higher than the MW values.
The similarity between M31 and the LMC is continued in the other panels that show correlations between \nhiav\ and $C_2$, $B_3$, and $C_4$.
The M31 points agree with the correlations seen by \citet{2024ApJ...970...51G} and strengthen the significance of these correlations.
It is quite surprising that the M31 UV extinction curves look LMC-like and the \nhiav\ ratios are also LMC-like.
Prior to this work, it was expected that the M31 extinction would look like the MW extinction given the overall solar metallicity of M31.
This may be an indication that the spatial limitation of MW UV dust extinction knowledge to be within 1~kpc of the Sun is significant and may not be applicable to the whole of the MW.


\subsection{Dependence on Galactocentric Distance}

A radial metallicity gradient in M31 has been inferred from a variety of methods, including the dust-to-gas ratio \citep{2014ApJ...780..172D}, H II regions \citep{2012ApJ...758..133S}, blue supergiants \citep{2022ApJ...932...29L}, and the tip of the Red Giant Branch (RGB) \citep{2025arXiv250418779L}. These and other studies show some evidence for a gradient in metallicity from Solar or super-Solar at the galaxy center dropping to sub-Solar values near the metallicity of the LMC at $R \sim 25$~kpc.
In all the methods, there is a large scatter in metallicity values at every distance from the center, and there is significant disagreement among the studies as to the nature of the suggested gradient. 


We do not see evidence for a metallicity gradient in the observed sample of sightlines.
The hydrogen column density has been measured along most of our sightlines using Ly$\alpha$ so the gas-to-dust ratio can be calculated since the total extinction, \av\ is also known.
There is no correlation between \nhiav\ with galactocentric distance for our sample.
Also, there is only a small range of values of $\nhiav\ = 0.5 \times 10^{22}$ to $1.4 \times 10^{22}$ atoms~cm$^{-2}$~mag$^{-1}$.
There is also a tight range of metallicities that we estimated from the UV spectra of the sample stars, $\log(Z) = -0.15$ to 0.25.
Using Fig.~6 of \cite{2014ApJ...780..172D}, the predicted metallicity at the location of sightlines with UV extinction, can be estimated.
There is no correlation between these predicted values and the values calculated for our sample sightlines.

There is also no correlation between the FM90 extinction parameters and galactocentric distance which should reflect the metallicity gradient in M31.
This result is consistent with our sample having a small range of metallicities. 
Despite the small range of \nhiav\ in our sample, Figure~\ref{fig_gdprops} shows possible correlations between \nhiav\ and the FM90 parameters especially $B_3$ and $x_o$.

Because the observed sightlines are toward young stars, they are biased toward regions in the spiral arms with recent star formation.
So, they may have different \nhiav\ values than undisturbed regions in M31 and not follow the metallicity/galactocentric distance relation for M31.

\subsection{Average Extinction Curves}

\begin{figure}[tbp]
\epsscale{1.15}
\plotone{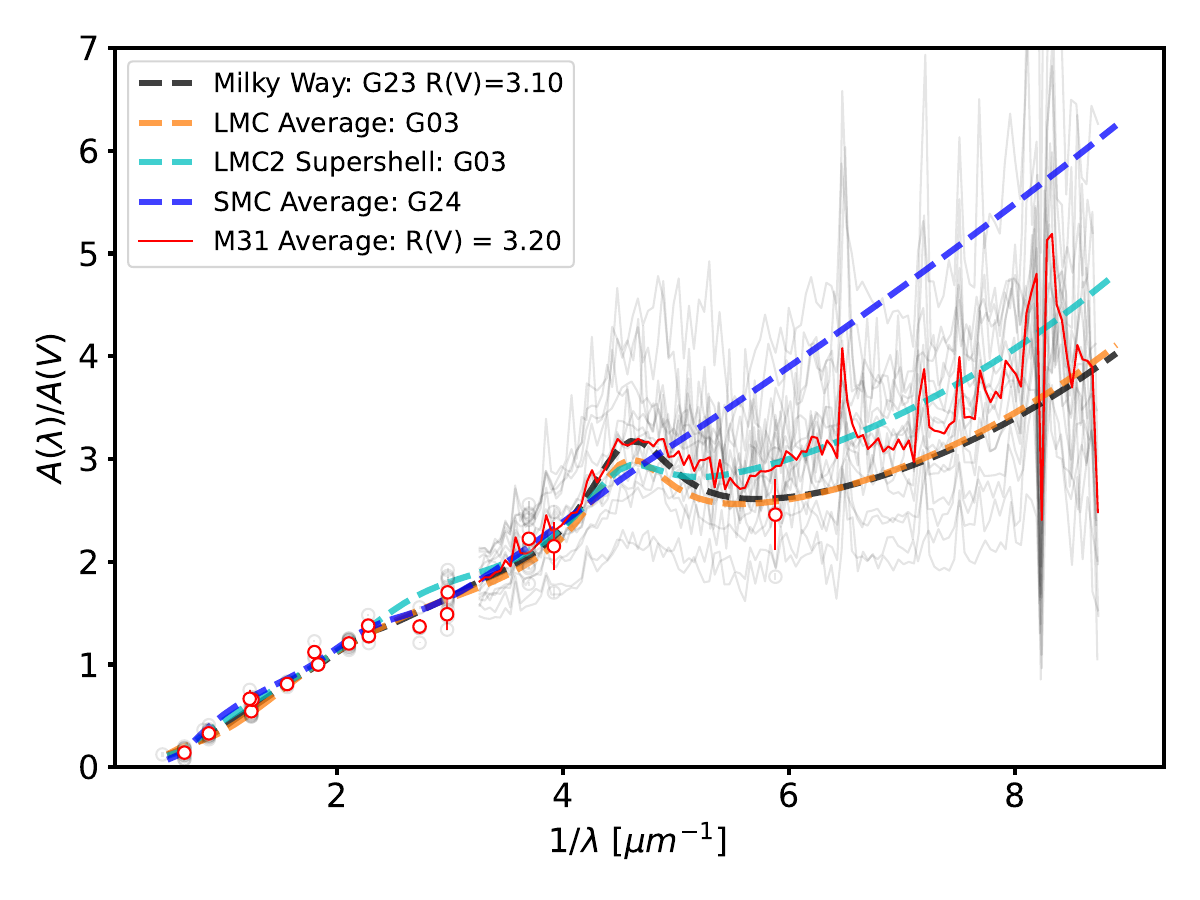}
\caption{The average extinction curve for M31 along with the measured individual extinction curves are shown.
For the photometric points, a minimum of 5 measurements is required for that band to be included.
For comparison, the extinction curves for the MW, the average LMC dust, the LMC dust near 30 Dor, and the SMC average dust are plotted.
\label{fig_aveext}}
\end{figure}

Figure~\ref{fig_aveext} shows the average M31 extinction curve, including all 17 sightlines, compared to the average curves of the MW, LMC2-30Dor, LMC-average, and SMC.
The M31 average $R(V) = 3.20$ and FM90 parameters are given in Table~\ref{tab_ext_fm90_params}.
The FM90 parameters were determined by fitting the average curve in \elvebv\ units with the same method as used by \cite{2024ApJ...970...51G}.

The average M31 extinction curve is very similar to the average LMC2-30Dor sightlines which lie close to the giant star-formation region.
In particular, it is very different from the MW even though these two galaxies have similar metallicities.
As can be seen from Figures~\ref{fig_fm90}-\ref{fig_aveext}, the FM90 parameters for most of the M31 sightlines lie close to those of the LMC2-30Dor sightlines.
The average M31 extinction curve is weaker around the bump, and steeper in the far-UV than the MW average but agrees well with the LMC2-30Dor average.
In particular, $C_2$, the slope of the UV extinction, $B_3$, the strength of the 2175~\AA\ bump, and $C_4$, the curvature of the extinction in the far-UV, are very similar for the M31 and LMC2-30Dor samples.
Both samples have $C_2$ values that are higher than the MW, weaker bumps than the MW as measured by $B_3$, and small values of $C_4$ indicating little curvature in the far-UV.
This can be seen in the average curves plotted in Figure~\ref{fig_aveext} and in the FM90 parameters plots in Figure~\ref{fig_fm90}.

\subsection{Environmental Dependence}

The differing extinction curves measured in the four Local Group galaxies, the MW, LMC, SMC and now M31 are evidence that the dust grains themselves are different from galaxy to galaxy.
Their properties may be dependent on a number of environmental parameters such as metallicity, star formation activity, and the gas-to-dust ratio which will affect the overall composition and size distribution of dust grains in each galaxy and in different regions of an individual galaxy \citep{1997ApJ...487..625G,2003ApJ...594..279G,2003ApJ...588..871C}.
It is not clear whether global characteristics such as galaxy metallicity or local conditions such as nearby star formation are the overriding factors for producing the observed dust grains along a line of sight.

Nevertheless, the correlations between the FM90 parameters in the MW, LMC, SMC, and now M31, indicate that as the 2175~\AA\ bump strength ($B_3$)
weakens, the FUV rise ($C_4$) also weakens and the UV slope ($C_2$)
strengthens \citep{2024ApJ...970...51G}.
This is a strong indication that there is a family of curves describing interstellar dust throughout the Local Group. 

Examining the {\it Galex} Near-UV and Far-UV images of M31, most or all of the sightlines in our sample lie in spiral arms near regions of active star formation \citep{2011Ap&SS.335...51B}.
The PHAT {\it HST} imaging of the seventeen individual stars shows most appear single and well separated, except for e5 and e9 which are clearly in stellar associations.



The steep, weak-bump sightlines in the SMC and the 30 Dor sightlines in the LMC have been associated with areas of active star formation and increased far-UV radiation and shocks that could affect the grain size distributions.
But, the recent discovery of additional sightlines in the SMC with significant bumps and less steep far-UV extinction suggests that self-shielding by dust clouds could explain how both type of sightlines could be present \citep{2024ApJ...970...51G}.
The weak bump and weak far-UV rise slope seen in M31, agree with the trends found in the bumpless SMC sightlines and the 30-Dor LMC sightlines. 
Similar sightlines are also found in some regions of the MW where strong UV radiation or shocks are present, in a sample of low-density sightlines \citep{2000ApJS..129..147C}, and toward the Trumpler~37 star cluster where the dust has been subject to supernova shocks \citep{2003ApJ...598..369V}.

\section{Summary}

This paper presents a large sample of UV extinction curves for M31 based on new spectra from {\it HST}/STIS.
This sample expands the number of reddened sightlines in M31 having UV spectra, including measured N(H I) from Ly$\alpha$, from four to seventeen.
All of the extinction curves have been determined using the pair method where the observed spectra of reddened stars are compared to unreddened stellar atmosphere models and then corrected for MW foreground extinction. 

This sample includes stars in M31 located from 5 to 16~kpc from the galactic center, but no correlation is found between the FM90 parameters and galactocentric distance which might be expected given the radial metallicity gradient in M31.
Examination of these new extinction curves reveals that the average extinction properties in M31 are different from those of the three other Local Group galaxies that have been previously studied, the MW, LMC, and SMC but quite similar to the average extinction curve in the 30-Dor region of the LMC.
The average M31 curve is weaker in the 2175~\AA\ bump and the far-UV curvature but steeper in the UV slope than the average MW curve.

Regional variations seen in the MW, LMC, and SMC show that different extinction characteristics can co-exist in areas with similar environmental pressures so the conditions along each sightline may be unique. These pressures may override global characteristics such as metallicity.




The code used for the analysis and plots is available\footnote{\url{https://github.com/karllark/hst_m31_ext}}\footnote{\url{https://github.com/karllark/measure_extinction}}\footnote{\url{https://github.com/karllark/extinction_ensemble_props}}
\citep{hstm31ext, measureextinction, extensembleprops}.
The uncalibrated HST STIS data used in this paper can be found in MAST: \dataset[10.17909/x0an-tw68]{http://dx.doi.org/10.17909/x0an-tw68}.
The custom reduced STIS spectra and measured extinction curves are available at \dataset[10.5281/zenodo.15849380]{https://doi.org/10.5281/zenodo.15849380}.
The M31 average extinction curve is available as the C25\_M31Avg average model in the dust\_extinction package\footnote{\url{https://github.com/karllark/dust_extinction}}  \citep{2024JOSS....9.7023G}.

\begin{acknowledgments}

This research is based on observations made with the NASA/ESA Hubble Space Telescope obtained from the Space Telescope Science Institute, which is operated by the Association of Universities for Research in Astronomy, Inc., under NASA contract NAS 5–26555. These observations are associated with GO program 14761. This work was supported by grant, HST-GO-14761.001-A. PYMJ acknowledges support by the Bulgarian Ministry of Education and Science under the program “Young Scientists and Postdoctoral Scholars”. MD acknowledges support from the Research Fellowship Program of the European Space Agency (ESA).
\end{acknowledgments}

\bibliography{everything2}
\end{document}